\documentclass[superscriptaddress, nofootinbib, prd]{revtex4}[12pt]
\usepackage{graphicx,bm}
\graphicspath{{figs/}}
\usepackage{amsmath,amssymb,amsfonts,rotating}
\usepackage{hyperref}
\usepackage[usenames,dvipsnames]{color}
\usepackage{mdwlist} 
\usepackage{slashed}
\usepackage{verbatim}
\usepackage{mathrsfs}
\usepackage[english]{babel}

\begin{document}

\def\cL{{\cal L}}
\def\be{\begin{eqnarray}}
\def\ee{\end{eqnarray}}
\def\bea{\begin{eqnarray}}
\def\eea{\end{eqnarray}}
\def\beq{\begin{eqnarray}}
\def\eeq{\end{eqnarray}}
\def\tr{{\rm tr}\, }
\def\nn{\nonumber\\}
\def\e{{\rm e}}


\title{Static spherically symmetric solutions in mimetic gravity: rotation curves \& wormholes}

\author{Ratbay Myrzakulov\footnote{Email address: rmyrzakulov@gmail.com}}
\affiliation{Department of General \& Theoretical Physics and Eurasian Center for
Theoretical Physics,\\Eurasian National University, 010008 Astana, Satpayev Str. 2, Kazakhstan}

\author{Lorenzo Sebastiani\footnote{Email address: lorenzo.sebastiani@unitn.it}}
\affiliation{Department of General \& Theoretical Physics and Eurasian Center for
Theoretical Physics,\\Eurasian National University, 010008 Astana, Satpayev Str. 2, Kazakhstan}

\author{Sunny Vagnozzi\footnote{Email address: sunny.vagnozzi@fysik.su.se}}
\affiliation{The Oskar Klein Centre for Cosmoparticle Physics, Department of Physics, Stockholm University, Albanova, SE-106 91 Stockholm, Sweden}
\affiliation{NORDITA (Nordic Institute for Theoretical Physics), KTH Royal Institute of Technology and Stockholm University, Roslagstullbacken 23, SE-106 91 Stockholm, Sweden}

\author{Sergio Zerbini\footnote{Email address: sergio.zerbini@unitn.it}}
\affiliation{Department of Physics, University of Trento,\\Via Sommarive 14, I-38123 Povo (TN), Italy}
\affiliation{Istituto Nazionale di Fisica Nucleare (INFN), Gruppo Collegato di Trento,\\Via Belenzani 12, I-38050 Povo (TN), Italy}
\affiliation{Trento Institute for Fundamental Physics and Applications (TIFPA),\\Via Sommarive 14, I-38123 Povo (TN), Italy}


\begin{abstract}
In this work, we analyse static spherically symmetric solutions in the framework of mimetic gravity, an extension of general relativity where the conformal degree of freedom of gravity is isolated in a covariant fashion. Here we extend previous works by considering in addition a potential for the mimetic field. An appropriate choice of such potential allows for the reconstruction of a number of interesting cosmological and astrophysical scenarios. We explicitly show how to reconstruct such a potential for a general static spherically symmetric space-time. A number of applications and scenarios are then explored, among which traversable wormholes. Finally, we analytically reconstruct potentials which leads to solutions to the equations of motion featuring polynomial corrections to the Schwarzschild spacetime. Accurate choices for such corrections could provide an explanation for the inferred flat rotation curves of spiral galaxies within the mimetic gravity framework, without the need for particle dark matter.
\end{abstract}


\maketitle

\section{Introduction}
\label{introduction}

The nature of the dark matter that appears to permeate the Universe, with an energy density content 5 times that of baryonic matter, is arguably one of the biggest open problems at the confluence of modern cosmology, astrophysics and particle physics. A wealth of evidence now exists in favour of its ubiquitous presence, which has been inferred from a variety of observations and considerations. These range from the rotation curves of spiral galaxies to the kinematics of galaxy clusters, from gravitational lensing to large-scale structure surveys and from the cosmic microwave background anisotropy spectrum to N-body simulations. In particular, it has been known since the 1970s thanks to the work by Rubin, Ford, Thonnard and Burstein \cite{rubin1,rubin2}, that the rotation curves of spiral galaxies appear to be asymptotically flat or even growing as one moves further from the center of the galaxy. That is, $v_{\text{rot}} \approx$ const or grows after reaching a maximum at approximately 5-10 kpc, far beyond the region where luminous matter is present. This is in stark contrast with the expected $v_{\text{rot}} \propto 1/\sqrt{r}$ expected on the basis of Keplerian mechanics, derived relying exclusively on the observed luminous matter content of galaxies. \\

The arguably simplest solution to this problem posits that spiral galaxies are embedded in diffuse halos of invisible dark matter, whose mass enclosed within a radius $r$ grows as $M(r) \propto r$. Several particle dark matter candidates have been proposed in the literature, and among them Weakly Interacting Massive Particles (WIMPs) no doubt stand out, their appeal arising both from the ''WIMP miracle'' (the observation that a thermal relic with weak scale annihilation cross-section naturally reproduces the correct dark matter abundance) and from the fact that such particles arise pervasively in many theoretically well motivated extensions of the Standard Model. Of course, a different approach to the problem of flat rotation curves is possible. That is, that the only matter content to be taken into account is the luminous one, whereas it is our theory of gravitation that requires revision (see e.g. \cite{nojiriodintsov1,nojiriodintsov2,nojiriodintsov3} for reviews). In this context, the example \textit{par excellence} is that of modified Newtonian dynamics (MOND), where a new effective gravitational force law is assumed \cite{mond}. This reduces to Newton's law at high acceleration, whereas its behaviour deviates at low acceleration. Attempts to include MOND within a complete theory exist, such as AQUAL \cite{aqual} and TeVeS \cite{teves}. Another notable attempt which is not far in spirit from MOND is that of metric skew tensor gravity (MSTG), see \cite{mstg}. Actually, even with GR and without particle dark matter, certain solutions with axial symmetry can reproduce the inferred rotation curves \cite{rein,vogt}. \\

Recently, a different approach to the missing mass problem in the context of modified theories of gravity, dubbed mimetic gravity, has been proposed \cite{m1}, the theory respecting conformal symmetry as an internal degree of freedom. This is achieved through the isolation of the conformal degree of freedom of gravity in a covariant fashion, by parametrizing the physical metric $g _{\mu \nu}$ in terms of an auxiliary metric $\tilde{g} _{\mu \nu}$ and a scalar field $\phi$, the mimetic field, as follows:
\begin{eqnarray}
g _{\mu \nu} = -\tilde{g} _{\mu \nu}\tilde{g} ^{\alpha \beta}\partial _{\alpha}\phi\partial _{\beta}\phi \ .
\label{mimetic}
\end{eqnarray}
The physical metric is invariant under conformal transformations of the auxiliary metric: $\tilde{g} _{\mu \nu} \rightarrow \Omega(t, {\bf x}) ^2\tilde{g} _{\mu \nu}$, $\Omega(t, {\bf x})$ being a function of the space-time coordinates. The equations of motion for the gravitational field can be obtained as usual by varying the action with respect to the physical metric, taking into account however its dependence on the auxiliary metric and the mimetic field. The corresponding equations of motion differ from Einstein's equations by the presence of an additional source term:
\begin{eqnarray}
G ^{\mu \nu} - T ^{\mu \nu} + (G - T)g ^{\mu \alpha}g ^{\nu \beta}\partial _{\alpha}\phi\partial _{\beta}\phi = 0 \ ,
\label{basic}
\end{eqnarray}
that is, the gradient of the mimetic field, $\partial _{\mu}\phi$, plays the role of 4-velocity of an additional perfect fluid with energy density $(T - G)$ and negligible pressure. For consistency, the following condition is to be satisfied:
\begin{eqnarray}
g ^{\mu \nu}\partial _{\mu}\phi\partial _{\nu}\phi = -1
\label{norm}
\end{eqnarray}
By taking the trace of Eq.(\ref{basic}) and using Eq.(\ref{norm}) we see that the trace equation will be satisfied even if $G \neq T$, that is, the gravitational field equations have non-trivial solutions and become dynamical even in the absence of matter. It is then argued that in a cosmological FRW setting this additional degree of freedom can mimic collisionless cold dark matter, hence the denomination of ``mimetic dark matter'' for the original model. It was further shown in \cite{m2} how the same equations of motion can be obtained by introducing a Lagrange multiplier term in the action, which constrains the norm of the gradient of the mimetic field. With the addition of a suitable potential for the mimetic field, the theory was shown capable to describe a variety of cosmological scenarios, including inflation, the current accelerating epoch, and bouncing solutions. \\

It was soon realized that mimetic gravity is a conformal extension of General Relativity, with dark matter arising from gauging out local Weyl invariance of the theory as an additional degree of freedom, which describes the flow of a pressureless fluid \cite{barvinsky}. Another early attempt to explain why the apparently innocuous parametrization of the metric in Eq.(\ref{mimetic}) yields different equations of motion was carried out in \cite{othermimetic1}, explaining this property in terms of variation of the action over a restricted class of functions, hence providing less conditions for the stationarity of the action and correspondingly more freedom in the dynamics.\footnote{This is a general property which occurs when one makes derivative substitutions into the action.} More recently still, it was realized that the appearance of an additional degree of freedom in mimetic gravity is directly related to singular disformal transformations \cite{othermimetic2,domenech,namba}.\footnote{See also \cite{ghalee,ghalee1,carvalho} for even more recent work on the role of disformal transformations in modified theories of gravity.} Recall that by virtue of diffeomorphism invariance of General Relativity, any metric $g _{\mu \nu}$ can be parametrized in terms of a fiducial metric $l _{\mu \nu}$ and a scalar field $\phi$ \cite{bekenstein}. Consider then a general disformal transformation of the form:
\begin{eqnarray}
g _{\mu \nu} \rightarrow \tilde{g} _{\mu \nu} = {\cal A}(\phi , X)g _{\mu \nu} + {\cal B}(\phi , X)\partial _{\mu}\phi\partial _{\nu}\phi \ ,
\end{eqnarray}
where $\phi$ is a scalar field and $X =  -g ^{\mu \nu}\partial _{\mu}\phi\partial _{\nu}\phi$. If the transformation is invertible, there is no additional degree of freedom associated to the scalar field and one recovers General Relativity. If, however, there exists the following particular relation between the conformal and disformal factors ${\cal A}$ and ${\cal B}$:
\begin{eqnarray}
{\cal B}(\phi , X) = \frac{1}{X}{\cal A}(\phi , X) - f(\phi) \ ,
\end{eqnarray} 
with $f(\phi)$ an arbitrary positive function of $\phi$ alone, then the resulting disformal transformation is singular and the number of degrees of freedom might change as a result of the transformation. This is at the origin of the extra degree of freedom describing dark matter in mimetic gravity. It has also been observed that the two approaches towards mimetic gravity and extensions such as Horndeski scalar-tensor mimetic theories, namely singular disformal transformations and Lagrange multiplier term in the action, are equivalent \cite{unipd}. \\

Interest in the mimetic gravity framework has spurred several follow-up works and extensions. Possible ghost instability issues and ways to avoid them have been addressed in \cite{barvinsky,othermimetic3,othermimetic4}. In \cite{modified,imperfect,initial}, extensions of mimetic gravity by the addition of higher derivative terms have been extensively studied (although this had already started in \cite{m2}), and it has been shown that these terms can alter the sound speed and lead to suppression of power on small scales, thus leading to an altered growth of structure which could potentially alleviate the ``small-scale problems'' faced by $\Lambda$CDM cosmology. Other interesting cosmological solutions to mimetic gravity have been studied in e.g. \cite{saadi,khalifeh}, and a detailed studied of cosmological perturbations within this framework has been carried out in \cite{Odm00}. Extensions to mimetic $F(R)$, $F(R,T)$ and Gauss-Bonnet gravity have been studied in \cite{OdMim,leon,DavoodMim2,epjc75,iran,viable,nurgissa,astashenok,oiko,odioiko,planckbicep}, while other extensions and connections to related theories of modified gravity such as the scalar Einstein-aether theory \cite{scalar,speranza1,speranza2,ali}, covariant renormalizable gravity \cite{covariant,field,early,carloni,blackhole,odintsov,witt,kluson,chaichian}, conformal teleparallel gravity, geometric scalar gravity, Horndeski gravity, unimodular gravity and other models have been explored in \cite{galileon,epjc75,nostrocomment,bufalo,silva,kan,guendelman,guendelman1,guendelman2,guendelman3,unipd,zerbini,hojman,ms,noui,
matarrese,vikman,cmsvz,living,unimodular1,unimodular2,langlois,zumalacarregui,weiner}. See also \cite{DavoodMim1,koshelev,nonlocal,inhomogeneous,saridakis} for generalizations of mimetic gravity. Static spherically symmetric solutions have been studied in \cite{miomimetic}, other black hole-like solutions in \cite{nordstrom}, the derivation of the Tolman-Oppenheimer-Volkoff equations to study compact objects and numerical solutions for quark stars and neutron stars has been carried out in \cite{stars,neutronquark}, whereas cylindrical solutions (cosmic strings) have been studied in \cite{cosmicstrings} and cosmological attractors in \cite{attractors}. \\

Currently, an open problem in mimetic gravity is whether or not it is possible to explain the inferred flat rotation curves of spiral galaxies. A possible solution has been proposed in \cite{epjc75}, consisting in a coupling between the gradient of the mimetic field and the matter hydrodynamic flux. However, such a coupling is phenomenologically problematic because of, e.g., local gravity tests. Here we will propose a different approach to the problem of rotation curves in mimetic gravity. To do so, we will extend the work on static spherically symmetric (SSS) solutions in mimetic gravity conducted in \cite{miomimetic}, and study other such solutions which could help address this problem. In particular, it has been shown that suitable corrections (linear and quadratic) to the Schwarzschild metric could provide a solution to the rotation curves problem without the need for particle dark matter. Importantly, it has been noted in \cite{miomimetic} that in the case of SSS solutions, the correspondence between the mimetic field and dark matter can only be formal, given that the mimetic field can assume imaginary values. In this work we will elucidate how, despite this only formal correspondence, it is possible in a SSS spacetime and on galactic scales to recover the phenomenology of dark matter. \\

The outline of the paper is as follows. In Section \ref{formalism} we will lay out the general formalism for mimetic gravity with the addition of a potential for the mimetic field. Section \ref{spherically} is devoted to considering static spherically symmetric solutions of the theory, and recovering the equations for reconstructing the potential for a given solution. We then demonstrate explicitly how to reconstruct the potential for some interesting cases, including a linear correction to the Schwarzschild metric and a traversable wormhole. Further applications of the constructed formalism are considered in Section \ref{rotation}, where we  complete the reconstruction procedure for scenarios which can solve the problem of rotation curves in mimetic gravity. We will discuss which metrics can exist as solutions in mimetic gravity, and where it will not be possible to obtain explicit solutions, we will carefully analyse limiting cases. We furthermore briefly discuss the planar motion of free-falling particles in the gravitational fields of the obtained solutions. We provide a brief summary and concluding remarks in Section \ref{conclusion}. \\

In the following, unless stated otherwise, we will set $k_{\mathrm{B}} = c = \hbar = 1$, while Newton's constant and the Planck mass are related by $8 \pi G_{N}\equiv M_{Pl}^2/8\pi=1$.

\section{General formalism for mimetic gravity with potential}
\label{formalism}

Let us start by considering mimetic gravity with the addition of a potential for the mimetic field. The action of the theory is given by:
\begin{eqnarray}
I = \frac{1}{2}\int_{\mathcal M} d^4x \ \sqrt{-g} \left [ R + \lambda \left ( g ^{\mu \nu}\partial _{\mu}\phi\partial _{\nu}\phi + 1 \right ) - V(\phi) \right ] \ ,
\label{action}
\end{eqnarray}
where $\mathcal M$ is the space-time manifold, $R$ is the Ricci scalar and $g$ is the determinant of the physical metric $g_{\mu\nu}$. Actions of the type in Eq.(\ref{action}) actually precede mimetic gravity, having been studied in earlier works, for instance \cite{lim,gao,capozziello}.\footnote{See also \cite{paliathanasis} for recent work on the topic.} The field $\phi$ represents the mimetic field~\cite{m1, m2}. As shown in \cite{m2}, a suitable choice for the potential $V(\phi)$ allows us to reproduce a huge variety of cosmological scenarios. The Lagrange multiplier $\lambda$ is used to constrain the norm of the gradient of the mimetic field. In fact, variation of the action with respect to $\lambda$ immediately provides us with the following constraint equation:
\begin{eqnarray}
g ^{\mu \nu}\partial _{\mu}\phi\partial _{\nu}\phi = -1 \,.
\label{vincolo}
\end{eqnarray}
From the above we see that, in a Friedmann-Robertson-Walker (FRW) spacetime, the mimetic field can be identified with cosmic time, up to an integration constant. \\

The equations of motion (EOMs) of the theory are obtained by varying the action with respect to the metric and the mimetic field. Variation with respect to the metric is carried out as follows:
\begin{eqnarray}
\frac{\delta I}{\delta g ^{\mu \nu}} & = & \int d ^4x \ \left [ \frac{\delta \sqrt{-g}}{\delta g ^{\mu \nu}}R + \sqrt{-g}\frac{\delta R}{\delta g ^{\mu \nu}} + \frac{\delta \sqrt{-g}}{\delta g ^{\mu \nu}}\lambda (g ^{\alpha \beta}\partial _{\alpha}\phi\partial _{\beta}\phi) + 1) + \sqrt{-g}\lambda \partial _{\mu}\phi\partial _{\nu}\phi 
\right.
\nonumber\\&&
\left.
+ \frac{1}{2}\sqrt{-g}g _{\mu \nu}V(\phi) \right ] = \nonumber \\
& = & \int d ^4x \ \left [ -\frac{1}{2}\sqrt{-g}g _{\mu \nu}R + \sqrt{-g}R _{\mu \nu} -\frac{1}{2}\sqrt{-g}g _{\mu \nu}\lambda (g ^{\alpha \beta}\partial _{\alpha}\phi\partial _{\beta}\phi + 1) + \sqrt{-g}\lambda \partial _{\mu}\phi\partial _{\nu}\phi
\right.
\nonumber\\&&
\left.
 + \frac{1}{2}\sqrt{-g}g _{\mu \nu}V(\phi) \right ] = 0 \ ,
\end{eqnarray}
where $R_{\mu\nu}$ is the Ricci tensor. Making use of Eq.(\ref{vincolo}) leads to the equation for the gravitational field:
\begin{eqnarray}
G _{\mu \nu} = -\lambda \partial _{\mu}\phi\partial _{\nu}\phi - \frac{1}{2}g _{\mu \nu}V(\phi) \ ,
\label{EOM1gen}
\end{eqnarray}
$G_{\mu\nu}:=R_{\mu\nu}-Rg_{\mu\nu}/2$ being the Einstein tensor. By taking the trace of Eq.(\ref{EOM1gen}) we obtain:
\begin{eqnarray}
G = \lambda - 2V(\phi) \implies \lambda = G + 2V(\phi) = -R + 2V(\phi) \ .
\label{lambda}
\end{eqnarray}
By inserting the above expression for $\lambda$ into Eq.(\ref{EOM1gen}), we get:
\begin{eqnarray}
G _{\mu \nu} = (R - 2V(\phi))\partial _{\mu}\phi\partial _{\nu}\phi - \frac{1}{2}g _{\mu \nu}V(\phi) \ .
\label{EOM1}
\end{eqnarray}
The above is an useful version of the EOMs associated with the gravitational field. One further EOM can be obtained if we vary the action with respect to the mimetic field $\phi$. The calculation is straightforward and yields:
\begin{eqnarray}
-\frac{1}{\sqrt{-g}}\partial _{\nu} \left ( \sqrt{-g}\lambda\partial ^{\nu}\phi \right ) = \frac{1}{2}\frac{\partial V}{\partial\phi} \,.
\label{fieldeq} 
\end{eqnarray}
Thus, Eqs.(\ref{EOM1},\ref{fieldeq}) represent the equations governing the dynamics of the system. On the other hand, by using Eq.(\ref{vincolo}), we find that Eq.(\ref{fieldeq}) is automatically satisfied. \\

A remark is in order here. When $2V=R$, that is, when the trace of the Einstein tensor is equal to the trace of the stress-energy tensor induced by the potential, one has $\lambda=0$ and recovers the field equations of General Relativity\footnote{In this case, the potential must be $V(\phi)=\text{const}$, namely a cosmological constant.}. On the other hand, when $\lambda\neq 0$, the theory exhibits new solutions. In \cite{m1} it has been shown that the degree of freedom introduced by the mimetic field can mimic the dynamics of cold dark matter in the limit $V=0$. We observe that we can interpret the contribution of the field in Eq.(\ref{fieldeq}) as being the contribution given by a fluid whose stress-energy tensor is given by:
\begin{eqnarray}
T_{\mu\nu}=-\lambda\partial_\mu\phi\partial_\nu\phi\equiv\rho u_\mu u_\nu \ .
\end{eqnarray}
The perfect fluid in question is pressureless, has energy density $\rho=-\lambda$ and 4-velocity $u_\mu=\partial_\mu\phi$, thanks to the fact that $\partial_\mu\phi\partial^\mu\phi=-1\equiv u_\mu u^\mu$. In an FRW spacetime such a 4-velocity is real [$u_\mu\equiv\partial_\mu\phi=(1,0,0,0)$], thus allowing the identification of the mimetic field with the degree of freedom corresponding to dark matter. On the other hand, on the SSS metrics which we will consider in our work, the fact that $\partial_\mu\phi$ must be a time-like vector leads to imaginary components and hence the association with dark matter can only be formal, as was first noted in \cite{miomimetic}. In other words, what is observable is the real quantity $\phi^2$, and the reconstruction of the potential $V(\phi)$ will be realized in a consistent way, as soon as one is dealing with real components of metric tensors. \\

\section{Static Spherically Symmetric space-times}
\label{spherically}

Let us continue by considering a Static Spherically Symmetric (SSS) space-time of the following form:
\begin{eqnarray}
ds^2=-a(r)^2 b(r) dt^2+\frac{dr^2}{b(r)}+r^2\left(d\theta^2+\sin^2\theta d\phi^2\right) \,,
\label{metric}
\end{eqnarray}
where $a(r),b(r)$ are functions of the radial coordinate $r$. The Ricci scalar is given by:
\begin{eqnarray}
\hspace{-1cm}R = -\frac{1}{r^2}\left[3r^2\,b'(r)\frac{a'(r)}{a(r)}+r^2 b''(r)+2r^2\,b\left(r\right)\frac{a''(r)}{a(r)}+4r\,b'(r)+4 r b(r)\,\frac{a'(r)}{a(r)}+2b(r)-2\right] \,,
\label{SSSRicci}
\end{eqnarray}
where the prime denotes a derivative with respect to $r$. From Eq.(\ref{vincolo}), assuming that the mimetic field depends on $r$ only (which itself follows from the symmetries of the EOMs) one has that:
\begin{eqnarray}
\phi'(r)=\sqrt{-\frac{1}{b(r)}} \ ,
\label{fieldSSS}
\end{eqnarray}
which leads to a pure imaginary expression for the mimetic field. The (0,0)- and (1,1)-components of the field equations read:
\begin{eqnarray}
1-b'(r)r-b(r)=\frac{V(\phi)r^2}{2} \ ,
\label{uno}
\end{eqnarray}
\begin{eqnarray}
\left(b'(r)r+2r \frac{a'(r)}{a(r)} b(r)+b(r)-1\right)=-\lambda b(r) r^2\phi'(r)^2-\frac{V(\phi)r^2}{2} \ .
\label{due}
\end{eqnarray}
By making use of Eq.(\ref{uno}) and taking into account Eq.(\ref{fieldSSS}), we can rewrite Eq.(\ref{due}) as:
\begin{eqnarray}
2a'(r) b(r)=\lambda a(r) r \ .
\label{duebis}
\end{eqnarray}
Finally, from Eq.(\ref{fieldeq}), one gets:
\begin{eqnarray}
\frac{d}{d r}\left(2a(r)b(r)\lambda r^2\phi'\right)=-a(r)r^2\frac{d V(\phi)}{d\phi} \ .
\label{fieldeqSSS}
\end{eqnarray}
In the above, $\lambda$ is given by Eq.(\ref{lambda}). Let us recall that Eq.(\ref{fieldeqSSS}) is a consequence of Eqs.(\ref{uno},\ref{due}), by virtue of the constraint on the mimetic field given by Eq.(\ref{fieldSSS}). In Appendix A we will also provide a more transparent method for deriving the SSS field equations of the model which relies on inserting the ansatz for the metric Eq.(\ref{metric}) directly into the action. \\

By taking Eq.(\ref{fieldSSS}) into account one readily obtains:
\begin{eqnarray}
\phi(r)=\pm i\int\frac{dr}{\sqrt{b(r)}}\,,\quad 4\frac{d}{dr}\left(a'(r) b(r)^{3/2}r\right)=a r^2\sqrt{b(r)}\frac{d V(r)}{d r} \ ,
\label{equations}
\end{eqnarray}
where we made use of the fact that the mimetic field $\phi$ is a function of $r$ and therefore, given an explicit solution for the metric, the potential can be treated as a function of $r$ also. These two equations, supplemented with Eq.(\ref{uno}), are the starting point for reconstructing the potential when a choice for $b(r)$ is made. In the following, we will also need to fix the form of $a(t)^2 b(r)$, which encodes the physical Newtonian potential. To do so, the equations will be rewritten keeping this aim in mind. Notice that when $\lambda=0$ and $V(\phi)=0$ we recover the equations of General Relativity with the Schwarzschild metric as the unique vacuum solution. On the other hand, when $V=2\lambda_0$ with $\lambda_0$ a cosmological constant, we recover the Schwarzschild-de Sitter solution.

\subsection{An example of a reconstruction procedure}

We now apply the formalism discussed thus far to the reconstruction of potentials which realize some simple choices of SSS metrics. To begin with, we note that for the choice
\begin{eqnarray}
b(r)=\left ( 1 - \frac{r_s}{r}\right) \ ,
\label{bex1}
\end{eqnarray}
where $r_s$ is a fixed radius, one has from Eq.(\ref{uno}) that:
\begin{eqnarray}
V(r)=0 \ ,
\end{eqnarray}
namely we are dealing with pure mimetic gravity. In this case, the second of Eqs.(\ref{equations}) leads to:
\begin{eqnarray}
a(r)=a_1+\frac{a_2}{\sqrt{1-\frac{r_s}{r}}}\left[\left(\sqrt{1-\frac{r_s}{r}}\right)\log\left[\sqrt{\frac{r}{r_0}}\left(1+\sqrt{1-\frac{r_s}{r}}\right)\right]-1\right] \ ,
\label{aex1}
\end{eqnarray}
where $a_{1,2}$ are dimensionless constants and $r_0$ is a radial scale. If $a_1=1$ and $a_2=0$ we recover the Schwarzschild solution of General Relativity [it corresponds to $\lambda=0$ in Eq.(\ref{lambda})], which is given by:
\begin{eqnarray}
ds^2=-\left(1-\frac{r_s}{r}\right)dt^2+\frac{dr^2}{\left(1-\frac{r_s}{r}\right)}+r ^2\left(d\theta^2+\sin^2\theta d\phi^2\right) \ .
\end{eqnarray}
On the other hand, when $a_2\neq 0$, we can set $a_1=0$. The metric then reads:
\begin{eqnarray}
\hspace{-1cm}ds^2=-a_2^2\left[\left(\sqrt{1-\frac{r_s}{r}}\right)\log\left[\sqrt{\frac{r}{r_0}}\left(1+\sqrt{1-\frac{r_s}{r}}\right)\right]-1\right]^2dt^2+\frac{dr ^2}{\left ( 1 - \frac{r_s}{r}  \right ) } + r ^2\left(d\theta^2+\sin^2\theta d\phi^2\right) \,.\nonumber\\
\end{eqnarray}
This result is in agreement with work undertaken in \cite{miomimetic}. There this kind of metric, its topological extensions and its physical implications have been studied in quite some detail (see also \cite{othermimetic2}).\\

Let us now consider a linear modification of the metric in Eq.(\ref{bex1}):
\begin{eqnarray}
b(r)=\left ( 1 - \frac{r_s}{r} + \gamma r \right ) \ ,
\label{bex2}
\end{eqnarray}
with $r_s,\gamma$ dimensional positive constants. From  Eq.(\ref{uno}) we get:
\begin{eqnarray}
V(r)=-\frac{4\gamma}{r} \ .
\end{eqnarray}
The solution for the mimetic field is then found to be an elliptic function. A closed expression can only be given in limiting cases. For $r \approx r_s$, one can neglect the linear correction in Eq.(\ref{bex2}) to obtain Eqs.(\ref{bex1},\ref{aex1}). That is, we can recover the Schwarzschild solution with the Newtonian term $r_s/r$. In this case, one gets:
\begin{eqnarray}
\phi(r\ll\sqrt{r_s/\gamma})=\pm i\left[r\sqrt{1-\frac{r_s}{r}}+\frac{r_s}{2}\log\left[2r\left(1+\sqrt{1-\frac{r_s}{r}}\right)-r_s\right]\right] \ .
\label{24}
\end{eqnarray}
From the above one immediately obtains:
\begin{eqnarray}
\phi(r\rightarrow r_s)\simeq \phi_s\pm2i\sqrt{r_s(r-r_s)}\,, \quad r\simeq r_s-\frac{(\phi_s-\phi)^2}{4r_s} \ ,
\label{25}
\end{eqnarray}
where:
\begin{eqnarray}
\phi_s=\pm\frac{i r_s}{2}\log(r_s) \ .
\label{26}
\end{eqnarray}
Reconstructing the potential leads to:
\begin{eqnarray}
V(\phi\rightarrow\phi_s)\simeq-\frac{4\gamma}{r_s}-\frac{\gamma(\phi_s-\phi)^2}{r_s^3} \ .
\label{Vex2}
\end{eqnarray}
On the other hand, for large distances, one can ignore the Newtonian term in Eq.(\ref{bex2}) and from the second in Eqs.(\ref{equations}) we obtain:
\begin{eqnarray}
a(\sqrt{r_s/\gamma}\ll r)=\frac{c_1(4+6\gamma r)+3c_2\sqrt{1+\gamma r}-c_2(2+3\gamma r)\arctan[\sqrt{1+\gamma r}]}
{\sqrt{1+\gamma r}} \ ,
\end{eqnarray}
with $c_{1,2}$ dimensional constants. If $c_1=1/4$ and $c_2=0$ the metric assumes the simple form:
\begin{eqnarray}
ds^2=-\left(1+\frac{3\gamma r}{2}\right)^2dt^2+\frac{dr^2}{(1+\gamma r)}+r ^2\left(d\theta^2+\sin^2\theta d\phi^2\right) \ .
\label{m22}
\end{eqnarray}
The mimetic field and the related potential are found to be:
\begin{eqnarray}
\phi(r)=\pm\frac{2i\sqrt{1+r\gamma}}{\gamma}\,,\quad r=-\frac{4+\gamma^2\phi^2}{4\gamma}\,,\quad V(\phi)=\frac{16\gamma^2}{4+\gamma^2\phi(r)^2} \ .
\label{Vex2bis}
\end{eqnarray}
\\
\indent
We note that the relation $4/\gamma^2<|\phi|^2$ must hold in order to guarantee the positivity of $r$. From Eqs.(\ref{Vex2},\ref{Vex2bis}) we can infer the behaviour of the potential by interpolating the given limits of the metric. The metric under consideration reduces to the usual Schwarzschild space-time for short distances, while at large distances linear and quadratic corrections in the coefficient $g_{00}(r)$ appear, as in Eq.(\ref{m22}). These corrections could have a physical relevance when considering the related Newtonian potential. To begin with, the quadratic correction is associated with a negative cosmological constant in the background and can be safely ignored when $\gamma^2 r^2$ is sufficiently small. On the other hand, the phenomenology of the linear term $\sim\gamma r$ could be intriguing in the context of the rotation curves of spiral galaxies (here we have assumed $\gamma > 0$). This term increases with distance and could help address the discrepancy between the inferred flat rotation curves of galaxies (whose form is usually attributed to collisionless cold particle dark matter) and the fall-off expected on the basis of Kepler's laws of motion applied to the luminous content of the galaxies, which we discussed in Section \ref{introduction}. In Section \ref{rotation} we will return to these kinds of metrics, reconstructing a solution where the cosmological constant-like contribution is independent from the one of the linear correction inside the Newtonian potential in view of the problem of the rotation curves of galaxies.\\

\subsection{Wormholes}

As a further example of reconstruction procedure, we consider the case where:
\begin{eqnarray}
b(r)=1-\frac{r_s^2}{r^2} \ ,
\end{eqnarray}
$r_s$ being once more a positive dimensional parameter. From Eq.(\ref{uno}) we immediately obtain:
\begin{eqnarray}
V(r)=-\frac{2r_s^2}{r^4} \ .
\end{eqnarray}
The second in Eqs.(\ref{equations}) fixes the metric, which is now given by:
\begin{eqnarray}
a(r)=\frac{\left[c_1+c_2\arctan\left[\frac{r}{\sqrt{r_s^2-r^2}}\right]\right]}{\sqrt{1-\frac{r_s^2}{r^2}}} \ .
\end{eqnarray}
In order to preserve the metric signature for $r_s<r$, we must set $c_2=0$. As a consequence, within the the choice $c_1=1$, the metric reads:
\begin{eqnarray}
ds^2=-dt^2+\frac{dr^2}{\left(1-\frac{r_s^2}{r^2}\right)}+r ^2\left(d\theta^2+\sin^2\theta d\phi^2\right) \ .
\label{mex3}
\end{eqnarray}
The mimetic field and the potential are found to be:
\begin{eqnarray}
\phi=\pm i r\sqrt{1-\frac{r_s^2}{r^2}}\,,\quad r=\sqrt{r_s^2-\phi^2}\,,\quad V(\phi)=-\frac{2r_s^2}{(r_s^2-\phi^2)^2} \ .
\end{eqnarray}
Note that the radial coordinate is real and positive. The metric given in Eq.(\ref{mex3}) is quite interesting, given that it can be employed to describe a (traversable) wormhole~\cite{visser}. A wormhole is a three-dimensional space equipped with two connected spherical holes: in a traversable wormhole, it is possible to enter the wormhole and exit in the external space again through a ``throat''  (see e.g. \cite{frolov, Jam, W, Furey:2004rq, bip, Staro, cond, mioworm,shaikh} for studies of wormholes in modified theories of gravity, and in particular \cite{varieschi} for a study of wormhole geometries in fourth-order conformal Weyl gravity, closely related to the work we are conducting). If $r_0=r_s$ is the radius of the throat, the metric in Eq.(\ref{mex3}) satisfies the requirements for traversable wormholes \cite{visser}, which we recall are given by:
\begin{enumerate}
\item $g_{00}(r)=1$ and $g_{11}^{-1}(r)=(1-r_s^2/r^2)$ are well defined for all $r \geq r_s$;
\item $g_{00}(r)=1$ is regular on the throat with 
$g_{00+}(r_s)=g_{00-}(r_s)=1$ and
$g_{00+}'(r_s)=g_{00-}'(r_s)=0$;
\item $g_{11}^{-1}(r_s)=0$ and $0<g_{11}^{-1}(r)$ for all $r_s<r$;
\item Given $\tilde b(r)=r_s^2/r$ such that $g_{11}(r)^{-1}=[1-\tilde b(r)/r]$, we have $\tilde b_+'(r_s)=\tilde b'_{-}(r_s)=-1<1$.
\end{enumerate}
In addition to the above, we further note that our space-time is asymptotically flat.

\section{Rotation curves in mimetic gravity}
\label{rotation}

Armed with the formalism thus far developed, we are ready to tackle the issue of rotation curves in mimetic gravity. Qualitatively, the underlying idea we will adopt is deceivingly simple, and is closely related to the winning idea in theories such as MOND or MSTG. If we can appropriately introduce a new scale in the theory (be it explicitly or dynamically), and relate this either to the scale where predictions of Newtonian gravity fail, or some preferred scale intrinsic to rotation curves data, we might be able to address the issue of rotation curves without having to postulate the presence of particle dark matter. We will mainly draw from two related examples in the literature: the first one is the static solution of conformal Weyl gravity \cite{Riegert}, successfully utilized in \cite{Mannheim,Mannheimrev,mannheimprl,mannheimlast,dwarf}. The second relevant example we will consider stems from certain solutions in $F(R) \propto R^n$ gravity \cite{capo}, successfully applied in \cite{salucci,salucci2,salucci3}.

\subsection{Linear and quadratic corrections}

We begin by considering the first example, inspired from the static solution of conformal Weyl gravity. In the Newtonian limit, two global corrections to the classical $\propto 1/r$ potential are considered: a linear one growing as $r$ and a quadratic one proportional to $r^2$. The quadratic correction appears with a negative sign, and thus represents a cosmological de Sitter background. It has been shown that these non-Newtonian corrections can explain the inferred shape of galactic rotation curves. Therefore we will attempt to reconstruct the appropriate mimetic potential which can lead to similar corrections to the Newtonian potential. For our purposes, it is convenient to make the following replacement:
\begin{eqnarray}
\tilde a(r)=a(r)^2 b(r)\,,
\end{eqnarray}
within the metric given by Eq.(\ref{metric}). In this way we have that:
\begin{eqnarray}
ds^2=-\tilde a(t)dt^2 + \frac{dr^2}{b(r)}+r^2\left(r^2 d\theta^2+\sin^2\theta d\phi^2\right) \ .
\end{eqnarray}
The following relations hold true:
\begin{eqnarray}
a(r)=\sqrt{\frac{\tilde a(r)}{b(r)}}\,,\quad a'(r)=\frac{1}{2\sqrt{\tilde a(r) b(r)}}\left(\tilde a'(r)-\tilde a (r)\frac{b'(r)}{b(r)}\right)\,.
\end{eqnarray}
Thus, evaluating the derivative of $V(r)$ by using Eq.(\ref{uno}), the second in Eqs.(\ref{equations}) reads:
\begin{eqnarray}
\frac{d}{dr}\left[\left(\tilde a'(r)b(r)-\tilde a(r)b'(r)\right)\frac{r}{\sqrt{\tilde a(r)}}\right]=
\sqrt{\tilde a(r)}\left[-b''(r)r-\frac{2}{r}\left(1-b(r)\right)\right]\,.\label{eqb}
\end{eqnarray}
Let us start by considering the following ansatz for $\tilde a(r)$:
\begin{eqnarray}
\tilde a(r)=1-\frac{r_s}{r}+\gamma _0 r-\lambda_0 r^2 \ ,
\label{aex}
\end{eqnarray}
with $r_s\,,\lambda_0\,,\gamma _0$ positive dimensional constants. That is, we are considering polynomial corrections to the Schwarszchild metric (with positive power), as previously advertised. The chosen form of the metric element $g_{00}(r)=-\tilde a(r)$ leads to a Newtonian term $r_s/r$ in the gravitational potential, which is expected to be dominant at small distances. On the other hand, the ``cosmological constant'' term $\lambda_0 r^2$ emerges on cosmological scales and has a de Sitter-like form (although, as explained in \cite{mannheimlast}, it is not associated with an explicit de Sitter geometry \textit{per se}). Let us for the moment ignore this term (which we will reintroduce later). The linear $\gamma _0 r$ term dominates at intermediate (galactic) scales, where the Newtonian potential $\Phi(r)$ reads:
\begin{eqnarray}
\Phi(r)=-\frac{(g_{00}(r)+1)}{2}\simeq-\frac{r_s}{2r}\left(1-\frac{\gamma _0 r^2}{r_s}\right) \ .
\label{Np}
\end{eqnarray}
From the above we see that, in addition to the classical Newtonian contribution, an extra term growing linearly with the distance $r$ appears. As is well explained in \cite{Mannheimrev}, the effect of such linear term in Eq.(\ref{Np}) is to lead, for sufficiently large $r$, to the following form for the rotation velocity profile $v(r)$ of galaxies:
\begin{eqnarray}
v ^2 \simeq v _{\text{Newt}} ^2 + \frac{\gamma _0 c^2r}{2} \ ,
\label{v2r}
\end{eqnarray}
where the speed of light $c$ has been reintroduced for the sole purpose of clarity and $v_{\text{Newt}}$ is the contribution expected from Newtonian mechanics alone (arising from the first term in Eq.(\ref{Np}). Thus, on sufficiently large scales, the rotational velocity does not fall-off as expected on the basis of Keplerian mechanics with only the luminous matter as source (thus requiring an extended dark matter halo to accommodate observations), but increases slightly as $\sqrt{r}$. This of course is particularly true for galaxies where the falling Newtonian contribution to Eq.(\ref{v2r}) cannot compete with the rising determined by the $\gamma_0$ term (depending of course on the size of $\gamma_0$), as is the case for small and medium sized low surface brightness (LSB) galaxies. This result is tantalizingly in agreement with astrophysical data for such galaxies, which exhibit a similar pattern (see e.g. discussion in \cite{mannheimlast} in the context of conformal Weyl gravity). In other words, the rotation curves of these galaxies start rising immediately. The next section of our work will be devoted to fixing the value of $\gamma _0$ using observational data from rotation curves of galaxies, together with an examination of the physical implications of our findings. \\

The situation is qualitatively different for sufficiently extended galaxies, such as some large, high surface brightness (HSB) galaxies. Firstly, for these galaxies the Newtonian contribution in Eq.(\ref{v2r}) might be sufficient to complete with the rising linear term, $\propto \gamma_0 r$. This would lead to a region of approximate flatness before any rise can begin, and once more this is consistent with the data for such galaxies (see e.g. \cite{mannheimlast}). Even more important is the fact that the effect of the de Sitter-like term can start to be important and has to be taken into consideration. Because of the negative sign with which it appears, its effect is to reduce the velocity, and thus Eq.(\ref{v2r}) is modified accordingly as follows:
\begin{eqnarray}
v ^2 \simeq v _{\text{Newt}} ^2 + \frac{\gamma _0 c ^2r}{2} - \lambda _0c ^2r ^2 \ .
\label{vquad}
\end{eqnarray}
Clearly, sufficiently far from the center of such galaxies, the quadratic term takes over and arrests the rising behaviour, $v (r) \propto \sqrt{r}$, caused by the linear term in Eq.(\ref{v2r}). Remarkably, this trend is precisely what is observed in high surface brightness (HSB) spiral galaxies, which are sufficiently large that the effect of the quadratic term can come to dominate in their outskirts. There is another important implication of the negative sign in front of the quadratic term in Eq.(\ref{vquad}). Given that $v^2$ cannot be negative, this implies that for scales greater than $R \sim \gamma _0/2\lambda _0$ bound orbits are no longer supported. In other words, this could provide a dynamical explanation for the maximum size of galaxies, determined by the interplay between the linear ($\gamma _0$) and the quadratic ($\lambda _0$) terms in Eq.(\ref{aex}). While this observation is intriguing and had already been made in conformal gravity, it is beyond the scope of our work to study its implications in more detail. \\

Let us now return to mimetic gravity, having discussed qualitatively the effect of the additional terms with respect to the Newtonian potential. We first reconstruct the form of the metric and then of the mimetic potential. From Eq.(\ref{eqb}) one finds:
\begin{eqnarray}
b(r)=\frac{\left(1-\frac{r_s}{r}+\gamma _0 r-\lambda_0 r^2\right)\left(1-\frac{3r_s}{r}+\frac{\gamma _0 r}{3}+\frac{c_0}{r^2}\right)}{\left(1-\frac{3 r_s}{2 r}+\frac{\gamma _0 r}{2}\right)^2} \ ,
\label{bex}
\end{eqnarray}
where $c_0$ is a constant. The form of the mimetic potential, which is found by making use Eq.(\ref{uno}), is quite involved and is given by:
\begin{eqnarray}
V(r)&=&-\frac{2}{3r^2(2r-3r_s+\gamma _0 r^2)^3}[54 r_s^2 r-27r_s^3+
171\gamma _0 r_s^2 r^2-8\gamma _0 ^2\lambda_0 r^7+r^4(16\gamma _0 +7r_s\gamma _0 ^2+324r_s\lambda_0)
\nonumber\\&&
+4 r_s r^3(-17\gamma _0-108r_s\lambda_0)+
r^6(\gamma _0 ^3-44\gamma _0\lambda_0)+6r^5(\gamma _0 ^2-12\lambda_0+12r_s\gamma _0\lambda_0)
\nonumber\\&&
-12c_0[-r_s+2r(1+r_s\gamma _0)+2r^3(\gamma _0 ^2+\lambda_0)-\gamma _0\lambda_0 r^4+3r^2(\gamma _0-3r_s\lambda_0)]] \ ,
\label{Vrot}
\end{eqnarray}
while the form of $\phi(r)$ must be inferred from the first in Eqs.(\ref{equations}) and can only be given implicitly. However, it is possible to study our metric analytically for some limiting cases. Let us set:
\begin{eqnarray}
c_0=\frac{9r_s^2}{4} \ ,
\label{c0}
\end{eqnarray}
for reasons which will become clear shortly. In fact, let us take $\gamma _0 = \lambda_0=0$ (appropriate for small distances), so that:
\begin{eqnarray}
\tilde a(r)=1-\frac{r_s}{r}\,,\quad b(r)=\frac{4(c_0+r(r-3r_s))(r-r_s)}{r(2r-3r_s)^2} \ .
\label{45}
\end{eqnarray}
We then see that it is convenient to set $c_0=9r_s^2/4$ as in Eq.(\ref{c0}) so that the following holds:
\begin{eqnarray}
b(r)=1-\frac{r_s}{r} \ .
\end{eqnarray}
That is, we recover the vacuum Schwarzschild solution of General Relativity. Analogously to what we did in Eqs(\ref{24},\ref{25},\ref{26}), we find:
\begin{eqnarray}
\phi(r\rightarrow r_s)\simeq \phi_s\pm2i\sqrt{r_s(r-r_s)}\,,\quad
r\simeq r_s-\frac{(\phi_s-\phi)^2}{4r_s} \ ,
\quad
\phi_s=\pm\frac{i r_s}{2}\log[r_s] \ .
\end{eqnarray}
Thus, in the limit where $\phi\rightarrow\phi_s$, the potential reads:
\begin{eqnarray}
V(\phi\rightarrow\phi_s)\simeq-\frac{32\gamma _0}{3r_s}+\frac{13\gamma _0(\phi-\phi_s)^2}{r_s^3} \ .
\end{eqnarray}
In Appendix B we will speculate on possible solutions when $c_0=0$ in Eq.(\ref{45}). \\

Let us now consider the case where $r_s=\gamma _0=c_0=0$, that is, at large (cosmological) distances:
\begin{eqnarray}
\tilde a(r)=b(r)=(1-\lambda_0 r^2) \ .
\label{50}
\end{eqnarray}
In this case we find precisely the static patch of the de Sitter solution. The mimetic field then takes the form:
\begin{eqnarray}
\phi\simeq\pm i\frac{\text{arcsin}\left[\sqrt{\lambda_0} r\right]}{\sqrt{\lambda_0}}\,,\quad
r=\pm\frac{\sin\left[\sqrt{\lambda_0}|\phi|\right]}{\sqrt{\lambda_0}}\,.
\end{eqnarray}
The positive values of $r$ belong to the range for which $0<r<1/\sqrt{\lambda_0}$, where $H_0^{-1}=1/\sqrt{\lambda_0}$ is the cosmological horizon of the de Sitter solution with positive cosmological constant. The potential is given by (with the correction in $\gamma _0$):
\begin{eqnarray}
V(\phi)\simeq 6\lambda_0\mp\frac{4\gamma _0}{3}\left(\frac{\sqrt{\lambda_0}}{\sin\left[\sqrt{\lambda_0}|\phi|\right]}+ 4\sqrt{\lambda_0}\sin\left[\sqrt{\lambda_0}|\phi|\right]\right)
\,.
\end{eqnarray}
This result (with $\gamma _0=0$) is consistent with \cite{m2}, where the de Sitter solution is investigated within an FRW universe. \\

Finally, let us take the case where $r_s=\lambda_0=c_0=0$, relevant to galactic scales, which implies:
\begin{eqnarray}
\tilde a(r)=\left(1+\gamma _0 r\right)\,,\quad
b(r)=\frac{4\left(1+\gamma _0 r\right)\left(3+\gamma _0 r\right)}{3\left(2+\gamma _0 r\right)^2}\,.
\end{eqnarray}
In this case, the potential given by Eq.(\ref{Vrot}) reads:
\begin{eqnarray}
V(r)=-\frac{2\gamma _0(16+6\gamma _0 r+\gamma _0^2 r^2)}{3r(2+\gamma _0 r)^3}\,.
\end{eqnarray}
The mimetic field is instead given by:
\begin{eqnarray}
\phi=\pm\frac{i}{2\gamma _0} \sqrt{3(3+4\gamma _0 r+\gamma _0 ^2 r^2)}\,,\quad
r=\frac{-6\mp\sqrt{9-12\gamma _0 ^2\phi^2}}{3\gamma _0}
\,,
\end{eqnarray}
where only the solutions with the positive sign inside $r$ yields a positive radius for $\gamma _0 > 0$. By making this choice, the potential can then be explicitly given:
\begin{eqnarray}
V(\phi) = -\frac{2\sqrt{3}\gamma _0 ^2\left(27-4\gamma _0 ^2\phi^2+2\sqrt{9-12\gamma _0 ^2\phi^2}\right)}{\left(3-4\gamma _0 ^2\phi^2\right)^{3/2}\left(-6+\sqrt{9-12\gamma _0 ^2\phi^2}\right)} \ .
\end{eqnarray}
Thus, it is possible to reconstruct the limiting cases of the potential $V(\phi)$, giving rise to the solution in Eqs.(\ref{aex},\ref{bex}) for the metric which leads to the Newtonian potential given by Eq.(\ref{Np}). In particular, recall that we have considered the limiting case of small distance (where we have recovered the Schwarzschild solution), cosmological distances (where we have found that the maximum distance achievable is, as expected, the de Sitter horizon) and, most intriguingly, on galactic scales. For the first time, we have shown that it is possible to reconstruct non-Newtonian linear and quadratic corrections to the Newtonian potential in mimetic gravity, which can address the problem of galactic rotation curves. Importantly, from the dependence of the potential on $\phi$, we can note that the potential is real, as is required for consistency.  \\

\subsection{Fit to rotation curves}

To complete our work, we need to utilize the data from rotation curves of galaxies to fix the values of the two free parameters in our solution, i.e., $\gamma_0$ and $\lambda_0$. Fortunately, the apparently laborious task is considerably simplified. In fact, given that we have essentially reproduced the conformal gravity potential in mimetic gravity, it is possible for us to adopt the results in \cite{mannheimlast,dwarf}, where the same parameters were fitted to rotation curves. The total sample fitted consists of 138 galaxies, 25 of them being dwarf galaxies. Furthermore, the sample is vast enough to include extended galaxies which are statistically sensitive to the de Sitter-like term $\lambda_0 r^2$. These 21 galaxies have data points which go sufficiently far from the optical disk region. For the full list of galaxies fitted, including references to the galactic databases, the reader is invited to consult the appendices of \cite{mannheimlast,dwarf} and references therein. \\

Adopting the analysis of\cite{mannheimlast,dwarf}, the end result is that the fit to the rotation curves through the potential given by Eq.(\ref{vquad}) is excellent, giving a reduced $\chi^2$ of $\chi ^2_{\text{red}} \simeq 1$.\footnote{Personal communication with the authors.} The corrections to the Newtonian potential capture the falling and growing features in the rotation curves quantitatively rather than simply qualitatively. Using such analysis, we determine the best-fit values to our $\gamma_0$ and $\lambda_0$ parameters in mimetic gravity [Eq.(\ref{aex})] to be \cite{mannheimlast,dwarf}:
\begin{eqnarray}
\gamma _0 \simeq 3.06 \times 10^{-30} \ {\rm cm}^{-1} \, , \quad \, \lambda _0 \simeq 9.54 \times 10^{-54} \ {\rm cm}^{-2} \, .
\end{eqnarray}
The size of $\lambda_0$ is best appreciated if expressed as $\sim (100 {\rm Mpc})^{-2}$, suggesting indeed that its most important influences are on scales of large galaxies or clusters. \\

Why is the fit to rotation curves so successful? Let us for the moment leave aside $\lambda_0$ (which is relevant only for the largest galaxies anyway), and recall that we introduced a single new scale in the theory, given by $\gamma_0$. Remarkably, the rotation curves data also possess an approximate universal scale. Considering the measured distance $R_{\text{last}}$ and rotational velocity $v_{\text{last}}$ of the outermost data in the rotation curves, we can form the combination $\gamma_{\text{last}} \equiv v_{\text{last}}/c^2R_{\text{last}}$. Within better than an order of magnitude (especially when focusing on LSB galaxies), the value of $\gamma_{\text{last}}$ for all the 138 galaxies are remarkably closely clustered around the best-fit value for $\gamma_0$. Thus, in a sense, the rotation curve data too possess a preferred scale, which has been identified with the  new scale we introduced through our non-Newtonian correction. In fact, this argument explains precisely why MOND and MSTG are so successful as well, at least with what concerns rotation curves. That is, they both possess an universal scale which is close to $\gamma_{\text{last}}$. In the case of MOND, this scale is approximately given by $a_0/c^2 \simeq 1.33 \times 10^{-29} \ {\rm cm}^{-1}$, whereas for MSTG this is approximately given by $G_0M_0/r_0c^2 \simeq 7.67 \times 10^{-29} \ {\rm cm}^-1$.

In mimetic gravity, the large-scale behaviour of the mimetic dark matter is only a ``geometrical effect". Our analysis reinforces this concept: even the small-scale phenomenology of rotation curves in mimetic gravity suggests that the invocation of extended particle dark matter halos in conventional $\Lambda$CDM might instead only be an attempt to describe such global geometrical effects in local terms. Further tests of such modifications, similarly to what happens in conformal gravity, could come from cluster scales or, even more intriguingly, dwarf satellite galaxies (for example by examining the gap between the luminous material in primary galaxies and their satellites). It is beyond the scope of our work to speculate further along these lines and we refer the reader to \cite{mannheimlast} for further discussions on the matter.

\subsection{An alternative solution: general power-law correction}

Instead of solely considering linear and quadratic non-Newtonian corrections, we could consider a more general case with a power-law correction (with arbitrary power):
\begin{eqnarray}
\tilde a(r)=1-\frac{r_s}{r}+\gamma r^m-\lambda_0 r^2 \ ,
\label{aex2}
\end{eqnarray}
with $m$ a positive real parameter. We can now repeat the same exercise we did previously, to reconstruct the metric and the mimetic potential. The metric is fully determined by Eq.(\ref{eqb}) to be:
\begin{eqnarray}
\hspace{-1cm}b(r)=\frac{\left(4(2+m)r^2-12(2+m) r_s r+9(2+m)r_s^2-4(m-2)\gamma r^{2+m}\right)(r-r_s+\gamma r^{1+m}-\lambda_0 r^3)}{(2+m)r\left(3r_s-2r+(m-2)\gamma r^{1+m}\right)^2}\,.
\nonumber\\
\label{bm}
\end{eqnarray}
Once again, we have set the integration constant in such a way as to recover the Schwarzschild solution Eq.(\ref{45}) when $\gamma=\lambda_0=0$, namely at short distances. At large distances, instead, we may set $r_s=\gamma=0$ and find the static patch of the de Sitter solution in Eq.(\ref{50}). At galactic scales, setting $r_s=\lambda_0=0$, we obtain:
\begin{eqnarray}
\tilde a(r)=1+\gamma^m r\,,\quad
b(r)=
\frac{4(1+\gamma r^m)(2+m+2\gamma r^m-m\gamma r^m)}{\left(2+m)((m-2)\gamma r^m-2\right)^2} \ .
\end{eqnarray}
The potential can be given implicitly by:
\begin{eqnarray}
V(r)=\frac{2m^2\gamma r^{m-2}\left(\gamma^2m^2r^{2m}+m(8-6\gamma r^m-4\gamma^2 r^{2m})+4(2+3\gamma r^m+\gamma^2 r^{2m})\right)}{(2+m)\left((m-2)\gamma r^m-2\right)^3} \ .
\label{vm}
\end{eqnarray}
A trivial case is that where $m=2$, for which we obtain again a de Sitter-like solution (but with a negative cosmological constant when $\gamma$ is positive):
\begin{eqnarray}
\tilde a(r)=b(r)=1+\gamma r^2 \ .
\end{eqnarray}
Here the field is given by $\phi= \pm i\text{arcsinh} [\sqrt{\gamma}r]/\sqrt{\gamma}$ and the potential by $V(\phi)\simeq-6\gamma$. \\

Why is a general power-law non-Newtonian correction to the gravitational potential interesting? Such a correction has recently been studied in the context of $F(R)$ gravity in \cite{capo}. In fact, in the low-energy limit of power-law $F(R) \propto R^n$ gravity, a $r^m$ correction to the Newtonian potential emerges, with $m$ related to the power $n$ as:
\begin{eqnarray}
m = \frac{12n^2-7n-1-\sqrt{36n^4+12n^3-83n^2+50n+1}}{6n^4-4n+2}
\end{eqnarray}
Considerations on the stability of the potential at large distances and constraints from Solar System tests dictate that $0 < m < 1$. In \cite{capo} the above model was fitted to 15 LSB galaxies. Once more, since we have reproduced such non-Newtonian potential in mimetic gravity, we can adopt the results from the corresponding fit. Although the sample is clearly limited with respect to the 138 galaxies studied in \cite{mannheimlast,dwarf}, an excellent fit is achieved, with $\chi ^2 _{\text{red}} \simeq 1$. The best-fit value occurs for $m = 0.817$ (corresponding to $n = 3.5$. Even more intriguing is a fit done to two objects in \cite{salucci} for which a particle dark matter explanation works fails to explain rotation curves successfully: the dwarf galaxy Orion and the low luminosity spiral NGC 3198 (which had not been analysed in \cite{capo}). Again, the value of the fit is excellent ($\chi ^2 _{\text{red}} \simeq 1$), whereas $\Lambda$CDM fails with these two objects. A further remark is in order here. Whereas the potential reconstructed in Eq.(\ref{vm}) has only been given in its implicit form, we have nonetheless shown which metrics with $\tilde{a}(r)$ given by Eq.(\ref{aex2}) can be reconstructed in mimetic gravity, by providing an explicit form for $b(r)$ in Eq.(\ref{bm}). Despite lacking an explicit form for the potential, the result is significant as it implies that the very accurate fits to rotation curves obtained in \cite{salucci} can be extended to mimetic gravity. Once more, we have shown how the problem of rotation curves can be successfully addressed within this framework.

\subsection{Planar motion}

In this subsection we will analyse the motion  of a free particle in the gravitational field of the solution represented by Eqs.(\ref{aex},\ref{bex}, \ref{c0}). We begin by considering the geodesic equation, which is given by:
\begin{eqnarray}
\frac{d^2 x^\mu}{d\tau^2}+\Gamma^{\mu}_{\alpha\beta}\frac{d x^\alpha}{d\tau}\frac{d x^\beta}{d\tau}=0\,,\quad \Gamma^\mu_{\alpha\beta}=\frac{g^{\mu\sigma}}{2}\left(\partial_\alpha g_{\sigma\beta}+\partial_\beta g_{\sigma\alpha}-\partial_\sigma g_{\alpha\beta}\right)\,,
\label{geo}
\end{eqnarray}
where $\tau$ denotes proper time and $\Gamma^\mu_{\alpha\beta}$ are the Christoffel symbols. Given that we are interested in planar motion, we can fix the coordinate $\theta$ as $\theta=\pi/2$ such that $d\theta/d\tau=0$. Thus, from the fourth component of Eq.(\ref{geo}) we have:
\begin{eqnarray}
\frac{d\phi}{d\tau}=\frac{l_0}{r^2}\,,
\end{eqnarray}
where $l_0$ is a constant and represents conserved angular momentum. From the first component of Eq.(\ref{geo}) we obtain:
\begin{eqnarray}
\frac{d t}{d\tau}=\frac{b_0}{\tilde a(r)}\,,
\end{eqnarray}
where $b_0$ is a constant and $\tilde a(r)$ is given by (\ref{aex}). For a time-like orbit with $ds^2/d\tau^2=-1$, we get:
\begin{eqnarray}
\left(\frac{d r}{d\tau}\right)^2=-b(r)\left(1-\frac{b_0^2}{\tilde a(r)}+\frac{l_0^2}{r^2}\right)\,,
\end{eqnarray}
where $b(r)$ is given by (\ref{bex}, \ref{c0}). From this equation we derive the following:
\begin{eqnarray}
\left(\frac{d r}{d\phi}\right)^2=-\frac{r^4 b(r)}{l_0^2}\left(1-\frac{b_0^2}{\tilde a(r)}+\frac{l_0^2}{r^2}\right)\,.
\end{eqnarray}
In the limit where $\gamma=\lambda_0=0$ we find that the following holds:
\begin{eqnarray}
\left(\frac{d u}{d\phi}\right)^2=\frac{(b_0^2-1)}{l_0^2}+\frac{r_s u}{l_0^2}-u^2+r_s u^3\,,
\quad u=\frac{1}{r}\,.
\end{eqnarray}
That is, we have recover the elliptic equation with $u^3$-correction to the Schwarzschild orbit. When $r_s=\lambda=0$, that is, on galactic scales, we get:
\begin{eqnarray}
\left(\frac{d u}{d\phi}\right)^2=\frac{4(3u+\gamma)}{3l_0^2(2u+\gamma)^2}(-u+b_0^2u-l_0^2u^3-l_0^2\gamma u^2-\gamma)\,,
\quad u=\frac{1}{r}\,.
\end{eqnarray}
Here, at distances $u\sim\gamma$ (i.e. $r\sim\gamma^{-1}$), the elliptic orbit is distorted by the $u^{-1}$-term.

\section{Conclusion}
\label{conclusion}

In the present work we have studied static spherically symmetric solutions within the framework of mimetic gravity. Building up on previous work in \cite{miomimetic}, we have shown that through the addition of a potential for the mimetic field we can reconstruct a large set of such solutions. Having provided the relevant reconstruction formulas, we have explicitly performed the operation for some simple choices of metrics. In doing so, we have shown how mimetic gravity admits traversable wormhole solutions. \\

Because of the form of the relevant equations, it had been recently realized that in a static spherically symmetric space-time the mimetic field assumes imaginary values, hence invalidating a direct connection with the degree of freedom associated to dark matter. A major part of this work has therefore been devoted to showing that, nonetheless, the behaviour of the mimetic field can be appropriately modified through an appropriate choice of potential, such that it can reproduce the desired dark matter phenomenology on galactic scales. To do so, we have reconstructed static spherically symmetric solutions which display linear, quadratic, and more generally power-law non-Newtonian corrections to the Newtonian potential. Such corrections have been identified elsewhere to provide a good fit to galactic rotation curves without need for particle dark matter. In particular, the linear correction introduces a new scale in the theory, which intriguingly can be identified with a preferred scale with the data possess too. We have explicitly shown for which choices of metrics can this procedure be carried out successfully, and where it has not been possible to obtain explicit an explicit expression for the potential, we have carefully studied limiting cases. \\

For the first time, we have shown that it is possible to explain the inferred shapes of the galactic rotation curves in mimetic gravity, without additional degrees of freedom, but relying solely on the addition of a potential for the mimetic field. Thus, mimetic gravity can provide a correct description of phenomena usually attributed to particle dark matter on small scales, with the addition of a single degree of freedom to General Relativity. In the future, more accurate studies of the topics we have explored are worthwhile conducting. It would be interesting to study the implications of such theory on cluster scales: in particular, the de Sitter-like non-Newtonian correction is expected to be most relevant on such scales. In particular, it is known that $\Lambda$CDM predicts an overabundance of structure on both small (dwarf galaxies) and large (clusters and superclusters) scales and it would be interesting to explore whether the $\lambda_0$ term could play a role in alleviating this discrepancy. We have also anticipated that further tests of the model we presented can be conducted on dwarf satellite galaxies, and we reserve the study of this proposal for future work.

\section*{Acknowledgements}

SV is supported by the Swedish Research Council (VR) through the Oskar Klein Centre. We are grateful to the anonymous referee for useful comments which greatly improved the quality of the work.

\section*{Appendix A \quad Alternative derivation of the equations of motion}

To derive the equations (\ref{uno}) and (\ref{duebis}) one may plugg the expression for the Ricci  scalar (\ref{SSSRicci}) into the action (\ref{action}) to obtain, after integration by part, the following Lagrangian (see \cite{mioLagr1})
\begin{eqnarray}
\mathcal L=2a\left(1-r b'(r)-b(r)\right)+a(r)r^2\lambda\left(b(r)\phi'(r)^2+1\right)-a(r)r^2V(\phi)\,.
\end{eqnarray}
Thus, the derivatives respect to $a(r)$ and $b(r)$ with (\ref{vincolo}) lead to (\ref{uno}) and (\ref{duebis}). The derivation respect to $\phi$ of the Lagrangian leads to (\ref{fieldeqSSS}).

\section*{Appendix B \quad Reconstruction of the potential for \boldmath{$c_0 = 0$}}

Let us take $\gamma=\lambda=0$ and $c_0=0$ in (\ref{aex}, \ref{bex}),
\begin{eqnarray}
\tilde a(r)=\left(1-\frac{r_s}{r}\right)\,,\quad
b(r)=\frac{4(r-3r_s)(r-r_s)}{(2r-3r_s)^2}\,.
\end{eqnarray}
In this case, the metric signature is preserved for $3r_s<r$. 
The potential reads
\begin{eqnarray}
V(\phi)=\frac{18r_s^2(r_s-2r)}{r^2(2r-3r_s)^3}\,.
\end{eqnarray}
The field results to be
\begin{eqnarray}
\phi(r)=\pm i \frac{\left[2r^2-8 r_s r+6 r_s^2+r_s\sqrt{r^2-4 r_s r+3 r_s^2}\log\left[r-2r_s+\sqrt{r^2-4 r_s r+3 r_s^2}\right]\right]}{2\sqrt{(2r-3r_s)(r^2-4r_s r+3 r_s^2)}}\,.
\end{eqnarray}
The potential can be explicitly reconstructed in some region of the space. If we take $r$ close to $3r_s$, we may write:
\begin{eqnarray}
\phi(r)\simeq\phi_0\pm i\sqrt{\frac{3}{2}}\sqrt{r-3r_s}\,,\quad
r\simeq\frac{1}{3}\left(9r_s+(\phi-\phi_0)^2\right)\,,\quad
\phi_0=\pm i\frac{\sqrt{r_s}\log[r_s]}{2\sqrt{3}}\,.
\end{eqnarray}
Finally the potential can be reconstructed as
\begin{eqnarray}
V(\phi\rightarrow\phi_0)\simeq-\frac{1458r_s^2(15r_s+2(\phi-\phi_0)^2)}{(9r_s+(\phi- \phi_0)^2)^2(9r_s+2(\phi-\phi_0)^2)^3}\,.
\end{eqnarray}

\end{document}